\documentclass[9pt,twocolumn,twoside]{pnas-new}
\usepackage{booktabs}
\usepackage{csvsimple}
\usepackage{jabbrv}

\templatetype{pnasresearcharticle} 

\begin{document}

\title{
Reassessment of Kepler's habitable zone Earth-like exoplanets with data-driven null-signal templates
}

\author[a,1]{Jakob Robnik}
\author[a,b]{Uro\v{s} Seljak}

\affil[a]{Department of Physics, University of California, Berkeley, CA 94720, USA}
\affil[b]{Lawrence Berkeley National Laboratory, 1 Cyclotron Road, Berkeley, CA 93720, USA}

\leadauthor{Robnik}

\significancestatement{
We propose a data-driven null-signal template (NST) method to reliably estimate the statistical significance of exoplanet transit candidates in the presence of unknown sources of signal contamination. NST accounts for all possible systematics in the data, and accurately reproduces false alarm distribution in the Kepler space telescope data, unlike currently established methods such as scrambling or inverting the light curves. We show NST to have a significant impact on the statistical significance of known habitable zone Earth-like candidates. Our analysis assigns False Alarm Probability of Kepler 452 b and KOI 2194.03 to be below 1\%, making Kepler 452 b potentially the closest rocky analog to Earth in the Kepler data in terms of period, instellation flux and stellar temperature.
}

\correspondingauthor{\textsuperscript{1} E-mail: jakob\_robnik@berkeley.edu}


\begin{abstract} 
One of the primary mission goals of the Kepler space telescope is to detect Earth-like terrestrial planets in the habitable zone around Sun-like stars. Unfortunately, such planets are at the detection limit. Estimating their statistical significance via false alarm probability (FAP) is crucial for their validation, and has a large impact on the estimate of their occurrence rate, which is of central importance for future spectroscopic missions searching for life signatures. Current methods estimate FAP by light curve inverting or scrambling, but we show that both of these approaches are unsatisfactory. Here we propose to modify the planet transit template by randomly shifting the transit times by small amounts. We show that the exoplanet search with the resulting Null Signal Template (NST) has the same statistical properties as with the true periodic template, which enables assigning a reliable star-specific FAP to every candidate. We show on simulations and on the real data that the method is robust to unmodeled noise contamination. We reevaluate the statistical significance of all 47 previously proposed habitable Earth-like and super Earth Kepler candidates and assign them star-specific NST based FAP. We identify 29 candidates with FAP below 1\%, 7 of whom are currently not considered confirmed. Among these are Kepler 452b with radius $1.5 R{\oplus}$, a period of 384 days, and KOI 2194.03 with radius $1.8 R{\oplus}$ and a period of 445 days, both around Sun-like G stars. Several well-known candidates should be considered marginal or likely false alarms, including Kepler 186f with 20\% FAP.
\end{abstract}

\dates{This manuscript was compiled on \today}
\doi{\url{www.pnas.org/cgi/doi/10.1073/pnas.XXXXXXXXXX}}

\maketitle
\thispagestyle{firststyle}
\ifthenelse{\boolean{shortarticle}}{\ifthenelse{\boolean{singlecolumn}}{\abscontentformatted}{\abscontent}}{}

\firstpage[11]{2}

\dropcap{K}epler space telescope discovered thousands of exoplanets by monitoring the brightness of over 150,000 stars with unprecedented precision \citep{borucki_kepler_2010}. Launched in 2009, the telescope was designed to detect Earth-sized planets in or near the habitable zones of Sun-like stars through the transit method—measuring periodic dips in stellar brightness caused by planets crossing in front of their host stars \citep{koch_kepler_2010}. The mission has revolutionized our understanding of planetary systems, revealing a diverse population of exoplanets, and suggesting that planets are common in the galaxy \citep{batalha_exploring_2014}.
Among Kepler’s key scientific goals was the identification of potentially habitable rocky planets—those exoplanets with small radii (typically below 1.8 Earth radius) receiving the right amount of stellar flux to support liquid water on their surfaces, which we will call habitable exoplanets. However, due to unexpectedly high stellar variability and other sources of noise and systematics \citep{gilliland_kepler_2011}, these planets remain close to or below the detection threshold. The number of habitable Earth-like candidates is small and their nature is controversial, which results in large uncertainty in the rocky habitable planets' occurence rate \citep{bryson_occurrence_2021}. The standard procedure of validating candidates through a ground-based radial velocity (RV) follow-up cannot be applied to these candidates due to the low signal-to-noise ratio of RV. One of the key ingredients of their validation is therefore statistical significance testing evaluating the false alarm probability (FAP) \citep{torres_validation_2015}, which is the focus of 
this paper.

The standard frequentist approach towards hypothesis testing is to first design a test statistic: a number that is high in the presence of the signal and low otherwise. Test statistic is then computed for the data at hand and compared against the distribution of the test statistic on the hypothetical null data where there is no signal, typically obtained by simulating the data generating process in the absence of a signal. The significance of the discovery is reported as the p-value: the probability of seeing a larger test statistic on the null data \citep{fisher_statistical_1955}. However, exoplanet surveys like the Kepler space telescope data \citep{KeplerSpaceTelescope} are too complex to allow reliable simulations. They contain stellar variability, non-Gaussian outliers, flares, rolling bands, blended background eclipsing binaries and other features \citep{jon_m_jenkins_kepler_2017, foremanplanetsearch, KeplerFlares, KeplerRollingBands}. Efforts in modeling the null hypothesis resulted in systematics removal algorithms \citep{KeplerErrorCorrection, mullally_identifying_2016, Stumpe2014, smith2012}, Gaussianization transformation for the outliers \citep{GMF}, procedures for computing the background false positive probability \citep{KeplerBackgroundFP, pastis, giacalone_vetting_2021} and Gaussian processes \citep{FourierGP, celerite} and matched filtering techniques \cite{jenkins2002, jenkins2010, kepler_process, robnik_matched_2021, ivashtenko_independent_2025} for the stellar variability. Modeling these sources of noise is beneficiary for the detection pipeline, as it can improve the test statistic. Nevertheless, 
the sources of noise are poorly understood, and their diversity and complexity can make a complete generative noise model intractable. These poorly understood sources of noise are an important source of false detections in the low signal regime \citep{mullally_keplers_2018}, and using simulations to quantify the false detection rate would underestimate the p-value. Machine learning classifiers \citep{shallue_identifying_2018,valizadegan_exominer_2022,exominerMultiplicityBoost} face the same issue, because they are only as good as the training set. Bayesian hypothesis testing is unreliable in this case, because the Bayes factor depends on the null likelihood which is even harder to construct than the null generative process and therefore in practice does not include important sources of false alarms, as for example in \cite{matesic_gaussian_2024}.

If a reliable generative model of the noise is not attainable, one alternative is to perturb the data to destroy the signal and then use it in place of the data. 
One such approach is bootstrap, i.e., to draw with replacement from the out-of-transit data to generate a null simulation. This approach was first used for the exoplanet transit data in \cite{castellano2000}. Later it was improved by taking into account correlated noise \citep{jenkins2002some, jon_m_jenkins_kepler_2017}. 
Finally, the state-of-the-art method for the Kepler data is block bootstrap, also called scrambling \citep{ScramblingProposal, thompson_planetary_2018, robovetter, CompletenessSteve}, where Kepler quarters are reshuffled, such that the real planets are no longer detectable. Inverting the data is also commonly used to complement scrambling \citep{CompletenessSteve}. Unfortunately, important sources of false alarms like outliers and flares are asymmetric with respect to the flux inversion, while data scrambling does not preserve the quasi-periodic artifacts. 
One particular troubling false alarm source are the so-called rolling bands. These temperature-dependent artifacts originate in focal plane electronics and show up as an increased noise level with non-trivial correlations on the transit-like timescales, thus can trigger transit search algorithms \citep{kolodziejczak_flagging_2010}. This is particularly severe in a relatively small number of readout channels. However, since a given Kepler target falls on a fixed sequence of channels over the course of a Kepler's orbital period of 372.57 days, it is periodically subject to the same instrumentation issues. The resulting false alarms therefore appear to lie in the Habitable zone for Sun-like stars \citep{jon_m_jenkins_kepler_2017}.
Scrambling reshuffles the data quarters or their blocks and therefore also significantly reduces the false alarm probability associated with the rolling bands, leading to incorrect estimates of FAP. As a workaround, Kepler team opted to construct both scrambled and inverted datasets, such that all sources of false alarms are represented \citep{thompson_planetary_2018}, but not in the correct proportions for outliers that are not symmetric under inversion (e.g. Figure \ref{fig: sim0}). This may have an impact on the downstream tasks, such as computing the reliability of the exoplanet detections and consequently their occurrence rate. It also prohibits the analysis of planets very close to the detection limit, where reliability becomes even more of an issue \citep{bryson_reliability_2020}.

Our proposed alternative is to modify the signal template rather than the data. 
For example, in gravitational wave detection, the detection time delay across different detectors is fixed by the speed of light. Therefore, if we search for a signal with a fixed, unphysically large delay between the detectors \citep{LigoNonCoincidential}, we can be confident to only find false detections, giving rise to  
the false detection distribution in the original search, so this null search can be used as a substitute for the null hypothesis simulation. 
A similar method has been recently proposed for periodograms, i.e. search for sinusoidal signal, such as Supermassive black hole binaries \citep{robnik_periodicity_2024}.
The idea is to construct a null signal template (NST) that is similar to the original template, so that it produces the same distribution of false alarms as the original template, while being sufficiently different that it does not trigger the true signal \citep{robnik_periodicity_2024}. Using the NST on the real data therefore acts as an effective null simulation.

In this work, we develop NST for the exoplanet transit search.
We construct it from the periodic template by randomly shifting the locations of the exoplanet transits by small time shifts. This ensures small overlap with the periodic template and preserves the false alarm distribution. This approach accounts for all possible systematics in the data. In the literature one 
differentiates between false alarm probability and false positive probability, which 
also addresses whether a given periodic event is an exoplanet versus alternatives such as eclipsing binaries of background stars. Here we rely on standard tools to 
establish this probability.

\section*{Null signal templates} \label{sec: nst}

\begin{figure}
    \centering
    \hspace*{-0.72cm}\includegraphics[scale = 0.27]{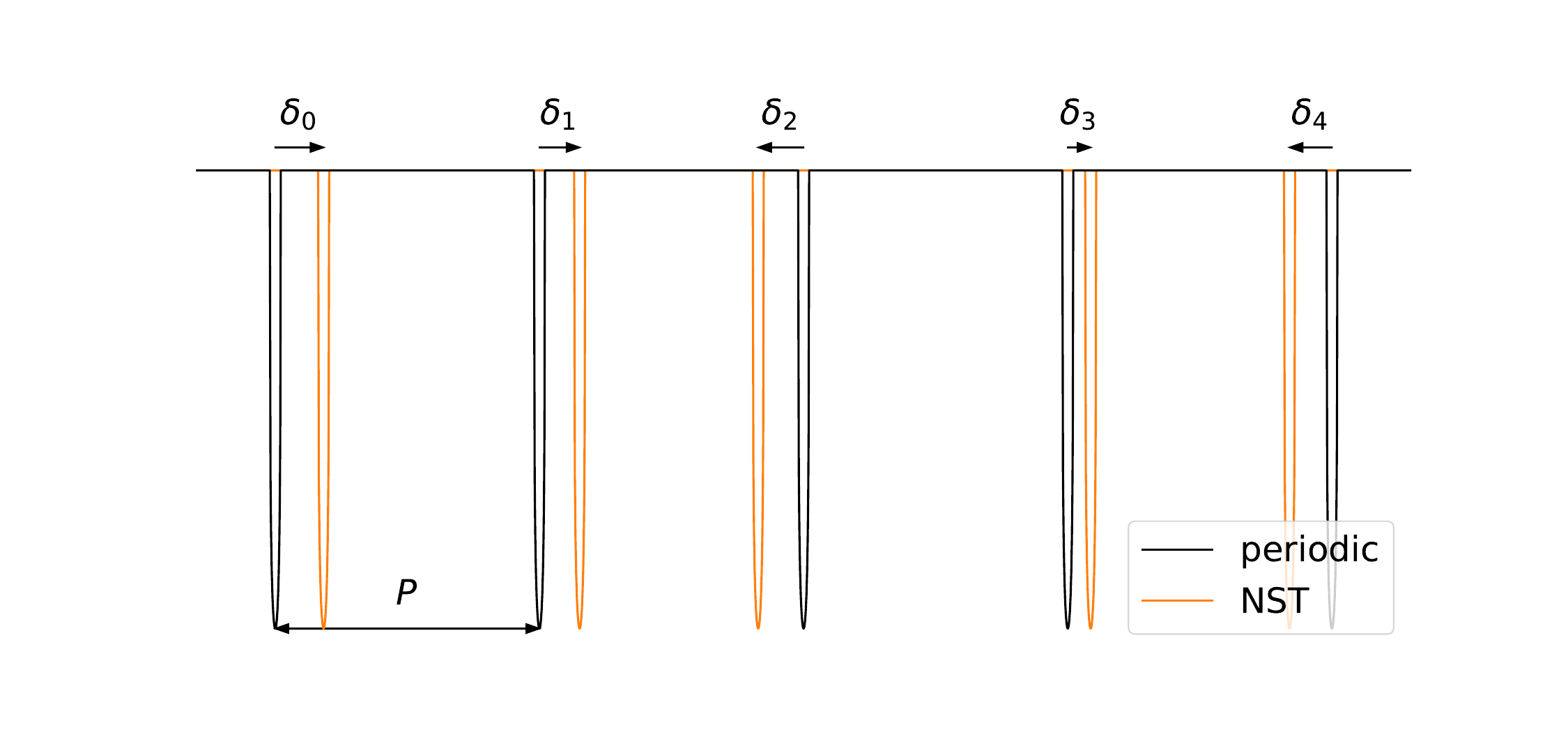}
    \caption{Null-signal template. The location of the $n$-th transit of the periodic template (black) is shifted by a random amount $\delta_n$ to obtain the null-signal template (orange).}
    \label{fig: template}
\end{figure}

\begin{figure*}
    \centering
    \includegraphics[scale=0.4]{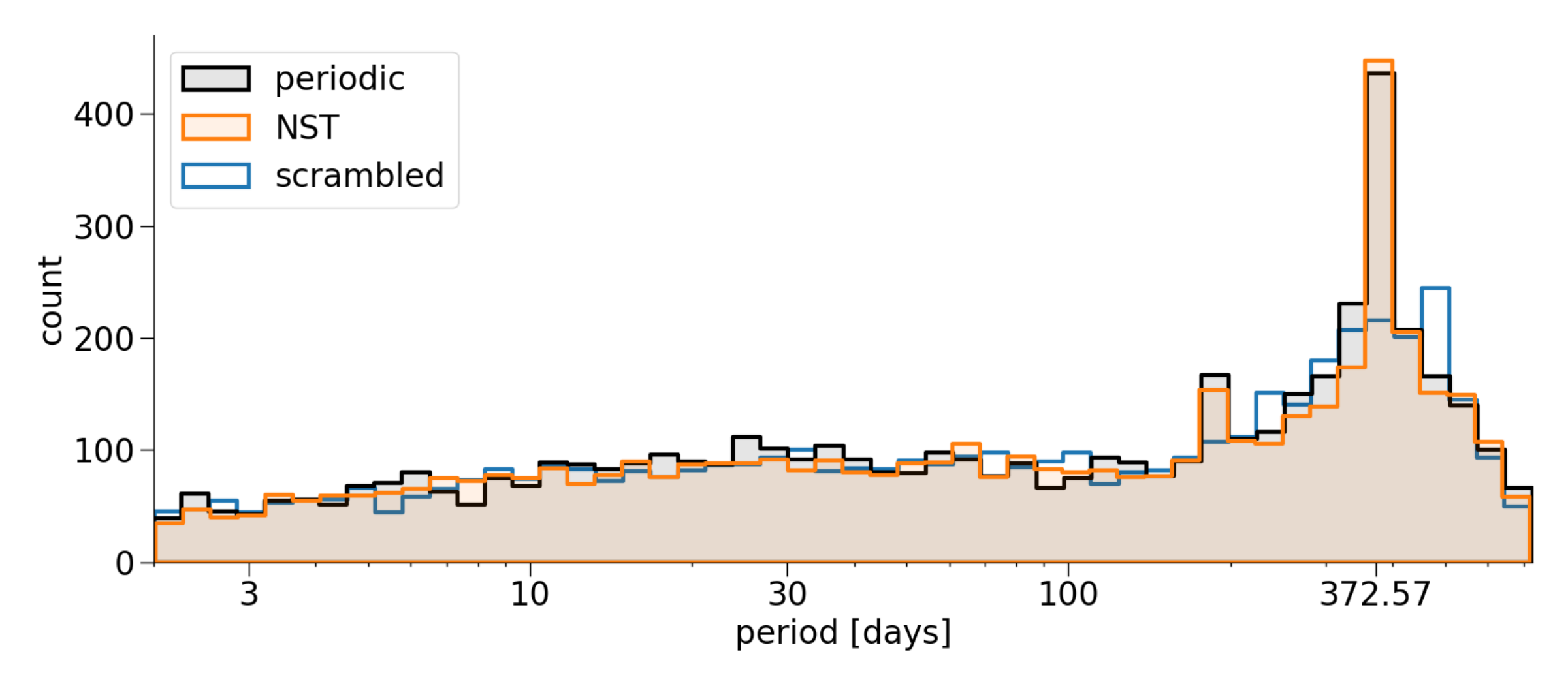}
    \caption{
    Period distribution of the exoplanet detections found in the inverted light curves of 5000 randomly chosen Kepler stars, which can be considered false alarms. The distributions with the periodic template (black) and NST (orange) are indistinguishable, up to statistical fluctuations. In particular, NST preserves the narrow peak of false alarms at the Kepler's orbital period of 372 days. In contrast, scrambling the data (blue) does not preserve this distribution and is therefore an unreliable null simulation, which may affect reliability estimations used in demographics analyses \citep{bryson_reliability_2020}.
    }
    \label{fig: inv}
\end{figure*}

Exoplanet transit search data X consists of flux time-series $\{x_i\}_{i = 1}^{N}$ at times $\{t_i\}_{i = 1}^{N}$. A planet signature is a periodic train of U-shaped transits, such that the $n$-th transit is centered at time $t_n = \phi + n P$, where $\phi \in [0, P)$ is the phase and $P$ is the period. A planet search algorithm \citep{jenkins_discovery_2015, GMF} varies the unknown period, phase and parameters of the transit shape and computes some test statistic $q(X)$ which is designed to maximize the contrast between the planet and no-planet hypothesis, for example the value of the signal-to-noise ratio at the optimal unknown parameters. Low p-value $P(q > q(X) \vert \text{no-planet})$ indicates a statistically significant discovery.
Ideally, one would simulate the no-planet hypothesis many times and compute the p-value as a fraction of simulations which yield $q$ higher than $q(X)$. Unfortunately, the no-planet hypothesis can be too complex to simulate. We here propose a family of the NSTs, such that the search with the NST on the original data can be used in place of the search with the original template on the null data. 

For a NST, $n$-th planet's transit is centered at time
\begin{equation} \label{eq: delta}
    t_n = \phi + n P + \delta_n.
\end{equation}
$\delta_n = 0$ corresponds to the perfectly periodic planet. The resulting template is shown in Figure \ref{fig: template}. 
We would like to make $\delta_n$ as small as possible, so that the NST is quasi-periodic, and yet not so small that it would directly overlap with the periodic template. Transits have no overlap if they are at least a transit duration apart. Transit duration of a circular orbit which is perfectly aligned with the line of sight is $\tau_K = (3 P / \pi^2 G \rho_*)^{1/3}$. Here $\rho_*$ is the density of the host star and $G$ is gravitational constant. Due to orbit misalignment, eccentricity and uncertainty in the stellar density, orbits are distributed around this value, but the probability of transit duration being significantly larger then $\tau_K$ quickly decays to zero. For example, for typical values of $10 \%$ uncertainty in the stellar parameters, eccentricity distribution as in \cite{KippingExomoon} and isotropically distributed orbit orientation, the probability of transit duration being larger than $2 \tau_K$ is $1.6 \times 10^{-3}$. We will therefore define the lower bound $\vert \delta_n \vert > a \tau_K$ with $a = 2$ as a requirement that NST template transits do not directly overlap with the periodic template transits. A possible complications would be transit timing variations \citep[TTV;][]{agol_detecting_2005,holman_use_2005}, a phenomenon where planet's transits do not occur completely periodically due to the gravitational interaction with other planets.
However, TTVs are typically very small, smaller than the transit duration, so the presence of TTVs does not affect this lower bound. Even in the rare cases of very strong gravitational interactions like that of Kepler 90 g-h \citep{YanTTV} it is very unlikely that the NST $\delta$ would match the real TTVs. Because of the small overlap between NST and the periodic template, the method can also be applied to stars with multiple planet candidates, which do not affect the results. We advise however that large confirmed planets first be removed. 

We also want to ensure that $\vert \delta_n \vert < P/2$, such that we eliminate degenerate situation where transits associated with different $n$ overlap with each other.
We therefore enforce the upper bound $\vert \delta_n \vert < b \tau_K$, with $b = (P_{\mathrm{min}} / 2) / \tau_K(P_{\mathrm{min}})$, where $P_{\mathrm{min}}$ is the smallest period considered in the search. This condition ensures that $\vert \delta_n \vert < P/2$ for all periods, because $\tau_K / P$ is a monotonically decreasing function of $P$.
For typical values of $P_{\mathrm{min}} = 2$ days and $\rho_{*} = \rho_{\odot}$, we get $b = 21.0$.

\begin{figure*}
    \centering
    \includegraphics[width=\linewidth]{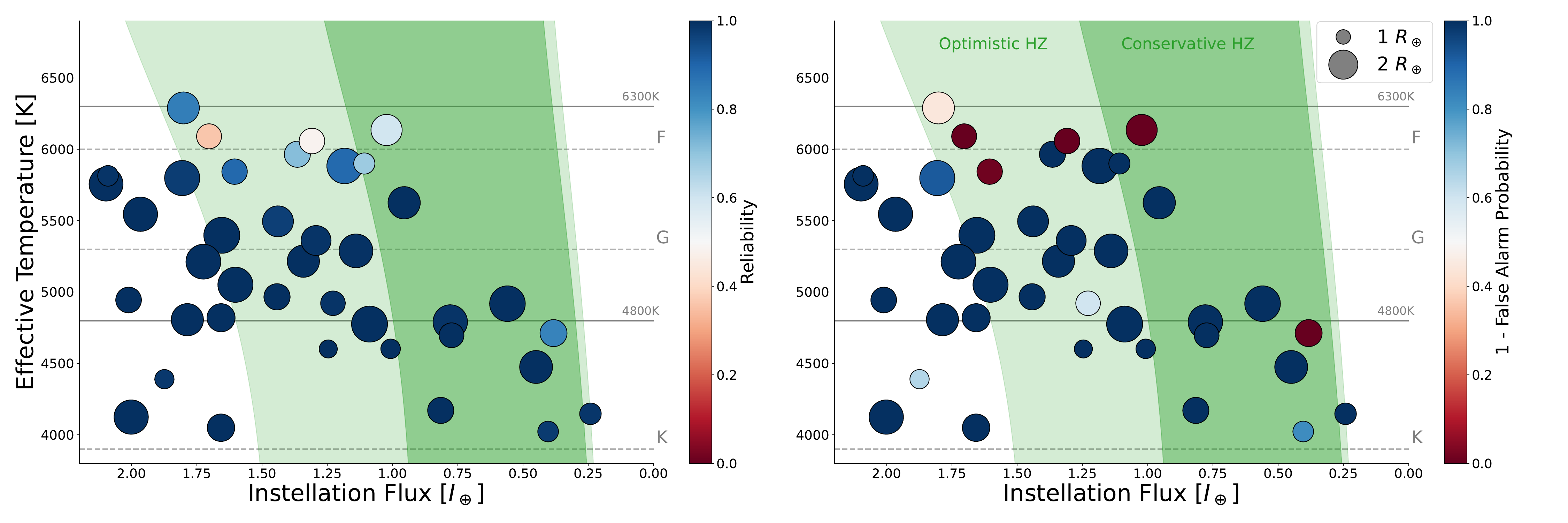}
    \caption{Earth-like and super Earth planet candidates from \citep{bryson_occurrence_2021} are shown in the effective temperature - instellation flux plane. Conservative and optimistic Habitable zones as defined in \citep{bryson_occurrence_2021} are shown as green regions. The size of the dots is proportional to the planet's radius. On the left, the candidates are colored by their reliability against false alarms from \citep{bryson_probabilistic_2020}. On the right they are colored by the false alarm probability as determined by NSTs. 
    Some of the candidates that were previously assigned a high reliability are no longer considered significant, and vice versa.
    }
    \label{fig: hz}
\end{figure*}

Taking this bounds into account we define
\begin{equation} \label{eq: nst delta}
    \delta_n = s_{n} [(1-u_n) a + u_n b] \tau_K(P),
\end{equation}
$s_n$ is the sign of $\delta_n$, i.e. positive (negative) $s_n$ indicates that the $n$-th transit is shifted to the right (left). We will have alternating shifts, $s_n = (-1)^n$, which ensures that in the case of only three transits, at most two NST transits can overlap with the periodic template. 
$0 < u_n < 1$ determine where in the range defined by the bounds is $\delta_n$. Each NST template will have fixed values of $u_n$ which are determined by a random independent draw from a uniform distribution. 
Different randomness realizations result in different NSTs which are approximately independent, and when applied to the given dataset each serves as one simulation of the null. This enables NST to 
assign p-values to 
individual events. 

Note that, just as inverting and scrambling the data, the NST method does not preserve strictly periodic astrophysical false positives, most notably eclipsing binaries (often in the background of the target star). Consequently, unlike the false alarm probability, the astrophysical false positive probability requires an additional analysis, and both are necessary for exoplanet validation.
Here we report VESPA results \citep{morton_vespa_2015, morton_false_2016} from \citep{CompletenessSteve, bryson_occurrence_2021}, which are based on false positive probabilities from Q1–Q17 DR25. However, VESPA is not intended for application to signals with Signal to Noise Ratio (SNR) below $10$ \citep{morton_false_2016}, which is the case for some of the candidates discussed in this work. Similarly, TRICERATOPS \citep{giacalone_vetting_2021} has not been demonstrated to be reliable for Kepler data, particularly at long orbital periods (greater than 50 days) and low SNR (below 15). We therefore do not report its values explicitly, although we verified that the reported probabilities are below 1\%.
Instead of differentiating between ${\rm SNR}<10$ and ${\rm SNR}>10$ (which would be validated according to VESPA criterion), we take 
a more conservative approach in this paper and 
we do not claim any of the candidates identified in this paper to be validated. 
A more reliable framework for estimating astrophysical false positive probabilities in the long period low SNR regime would need to be developed, which lies beyond the scope of this work.

\section*{Testing NST on inverted Kepler data}

Validation of NST on various simulated 
datasets is presented in Materials and Methods section below. 
To validate the NST method further we apply it to the inverted Kepler data, which serve as very realistic null simulations, containing various artifacts, such as rolling bands, which are approximately symmetric to inversion. 
We show that distribution of false alarms obtained with NST on this data exactly matches the true false alarm distribution, while scrambling of these data does not properly capture the rolling bands.

We take PDCSAP flux \citep{jenkins2010initial, jon_m_jenkins_kepler_2017} of 5000 randomly chosen Kepler stars. We first mask the transits of the known planets and then invert the dataset to eliminate the impact of any potentially unknown planets. 
We perform the exoplanet transit search with the pipeline developed in \citep{robnik_matched_2021} and only keep the exoplanet detections with Bayes factor larger than 1.

Figure \ref{fig: inv} shows the distribution of the exoplanet detection periods, similarly to Figure 2 in \citep{thompson_planetary_2018}. 
The analysis was performed both with the periodic template (black) and the NST (orange). Both yield the same period distribution, including the sharp peak at the Kepler's orbital period, which is typical of the rolling bands induced false alarms. The analysis with the periodic template on the scrambled data (blue) however does not match the unscrambled results. Notably, the peak at Kepler's orbital period is significantly broadened. 
This is a very significant result, as it demonstrates that NST gives more reliable null simulations than the current state-of-the-art scrambling of the data. This can have significant implications both for the exoplanet occurrence rate studies and for the validation of the individual exoplanet detections.

\section*{Results} \label{sec: results}

As an application of the method, we reevaluate
the statistical significance of several Earth-like exoplanet candidates. We take the list from \cite{bryson_occurrence_2021}. The candidates on the list are found among the $68,885$ Kepler's stars satisfying conditions in \citep{bryson_occurrence_2021}, mainly that the Gaia data for the star is available, majority of the Kepler's data time span is available, the star is on the main sequence and has effective temperature in the range $3900 K - 6300 K$. For planets to be considered Earth-like or super Earth planet candidates they must have radius in the range $0.5 R_{\oplus} - 2.5 R_{\oplus}$ 
Note that planets are unlikely to be rocky above $1.8 R_{\oplus}$.
We consider planets with instellation flux in the range  $0.2 I_{\oplus} - 2.2 I_{\oplus}$ which encompasses the habitable zone. The resulting 54 candidates are shown in Figure \ref{fig: hz}.

\begin{figure}
    \centering
    \hspace*{-0.5cm}\includegraphics[scale=0.49]{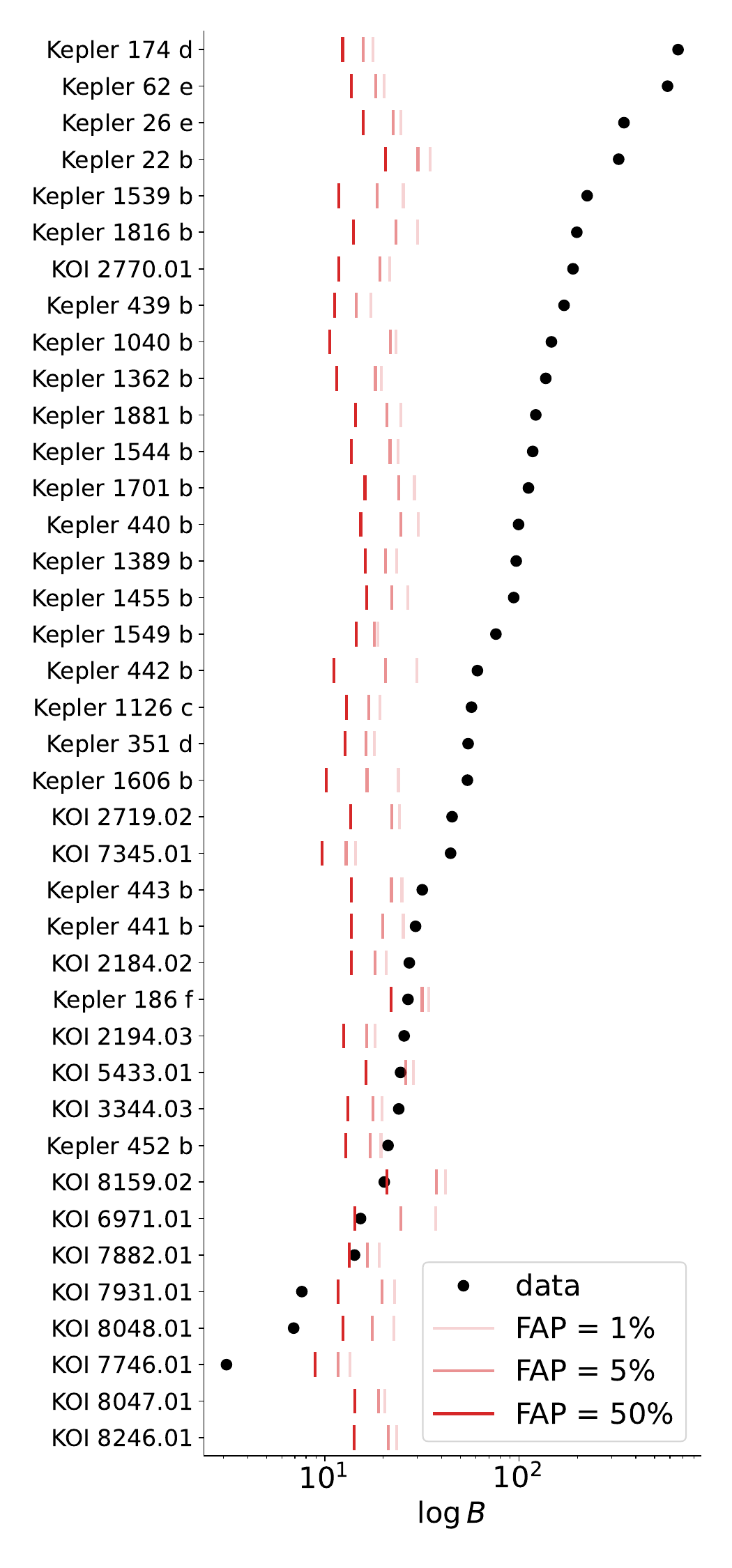}
    \caption{
    Reevaluation of Earth-like and super Earth planet candidates from \citep{bryson_occurrence_2021}. Logarithm of the Bayes factor for candidates is shown with a black dot. The null distribution quantiles are obtained with the NST method and are computed for each star separately, because different stars might have different noise properties. The red lines denote the quantiles, e.g. 1\% of the NST realizations produced a candidate with a Bayes factor surpassing the 1\% quantile line. The candidates surpasing the 1\% quantile can be claimed as a 1\% p-value discovery, which under the adopted standard should be treated as validated if they also have sufficiently low astrophysical false positive probability. 
    Candidates with CONFIRMED disposition in the Kepler catalog have Kepler names, the others are given the Kepler object of interest (KOI) identification number. The last two candidates have negative $\log B$ and are thus not shown on displayed plot.
    We note that NST quantiles vary 
    with Bayes factor $B$ from object to object, showing that individual star NST analysis is required for a reliable FAP of individual candidates. 
    }
    \label{fig: quantiles}
\end{figure}

For each system containing a candidate we first perform the exoplanet search to independently identify the proposed candidates. We reanalyze them and compute their Bayes factors $B$ and Signal to Noise Ratio (SNR), as described in \citep{robnik_statistical_2022}.
We find comparable results between $B$ and SNR, and we will 
focus on $B$ in the following, 
but we also report SNR results in Table \ref{table}. 
Seven of these candidates, KOI 
5276.01, 7894.01, 7915.01, 8107.01, 8242.01, 8033.01,  8063.01, 
were flagged as unreliable by the pipeline for various reasons, for example because the candidate is no longer significant after the sudden pixel sensitivity drops have been identified and removed from the data. 
We therefore do not consider these candidates in this work. We also do not consider the candidates which have the probability of being an astrophysical false positive as computed in \citep{morton_false_2016} larger than $20\%$, leaving us with 39 candidates. 
After having analyzed the candidates, we eliminate them from the time-series and search for additional planets with the NSTs. We repeat the analysis with 300 realizations of the NSTs, thus effectively obtaining 300 null simulations. In Figure \ref{fig: quantiles} we report the null hypothesis quantiles for the Bayes factor test statistic.
Since the candidates were found among $68,885$ stars but we have only performed null simulations on specific stars, we add the penalty $\log 68,885$ for the multiplicity of trials to the $\log B$ null results \citep{bayer_look-elsewhere_2020, robnik_statistical_2022}.
Our results are presented 
in Table \ref{table}
and Figure \ref{fig: quantiles}.
We observe that NST quantiles vary 
with Bayes factor $B$ and SNR from object to object. This 
is because $B$ and SNR
as test statistics can be affected by the stellar variability and systematics such as rolling bands, which vary from star to 
star. This 
demonstrates that individual star NST analysis is required for a reliable FAP of individual candidates. 

We identify 29 candidates 
having FAP p-value below 
1\%. Among these
are 7 without 
previously being 
assigned such status by 
Kepler. Notably, KOI 2194.03 
is in the optimistic habitable zone with a Sun-like star and radius $1.8 R_{\oplus}$.
Of interest is also Kepler 452 b which is very Earth-like, having radius $1.46 R_{\oplus}$, period 385 days, instellation flux $1.1 I_{\oplus}$ and a star with an effective temperature 5900 K. 
It has been validated by \cite{jenkins_452b, morton_false_2016}, but it is considered controversial \citep{burke_re-evaluating_2019} on the basis that its statistical reliability against false alarms is not below $1\%$.
However, these estimates are based on an average false alarms rate across all Kepler stars, making them less reliable than the star specific FAP. Star-specific NST analysis in this paper gives it a false alarm probability below $1\%$, so our work supports its validation as defined in \cite{jenkins_452b}.

Table \ref{table} shows VESPA \citep{morton_false_2016} eclipsing binary probabilities from \citep{bryson_occurrence_2021, bryson_probabilistic_2020}.
We have also confirmed that TRICERATOPS \citep{giacalone_vetting_2021} assigns the newly identified planets an astrophysical false positive probability below $1 \%$. In these calculations we have included a prior odds of 10 that was ignored in \citep{giacalone_vetting_2021}. We note however that neither VESPA nor TRICERATOPS were calibrated on the Kepler data in terms of giving correct empirical coverage and were not designed to work at signal-to-noise ratios considered in this work.

Bottom eight candidates in Table \ref{table} and Figure \ref{fig: quantiles} have FAP of order 50\% or larger, and are thus likely false alarms. 
The remaining candidates have FAP ranging from 1\% to 50\%, and should thus not be considered validated. Among them is the well-known Earth-like exoplanet candidate Kepler 186 f  with 20\% FAP.


\section*{Discussion}

We have established NST as a reliable method for producing effective null simulations in the exoplanet transit search on realistic datasets like the Kepler data. Both on simulations and on real data it produces correct null distributions, while inversion and scrambling of the data often fail at this task. This could have significant implications on demographic studies, specially for the long-period small-radius candidates in the habitable zone around G stars. 
NST method is also straightforwardly applicable to other exoplanet surveys, such as TESS data \citep{ricker_transiting_2015}, where scrambling is not used.
We used the NST method to reevaluate previously proposed candidates close to the detection limit and  assigned them individual FAP. Our analysis designates several candidates as being likely false alarms, and several more are being given FAP above 1\%, which is 
considered a validation threshold. 
On the other hand, we identify some new planet candidates like KOI 2194.03, which was previously considered non-validated and Kepler 452 b which was considered controversial, due to the low ensemble reliability against false alarms. 
This makes 452 b the closest Earth analog in the Kepler data in terms of radius, period, instellation flux and stellar temperature. 
Other identified (p-value $<$ 0.01\%) and potentially rocky ($R < 1.8 R_{\oplus}$) planets in the conservative habitable zone or close to the conservative habitable zone are Kepler 1544 b, 440 b, 441 b and 442 b.

Our FAP analysis is more reliable than population-level analysis because it is evaluated on a star by star basis, not as an average over all stars: stars have very different individual  properties such as stellar variability, as well as systematics such as rolling bands, leading to more or less false alarms at the same Bayes Factor $B$ or SNR. We 
show clear evidence of this in 
Figure \ref{fig: quantiles} and Table \ref{table}, where NST quantiles vary with Bayes factor and SNR. Since we perform 300 NST analyses per star 
we are able to assign reliable FAP down to 1\%, which is considered a validation threshold.
Furthermore, NST is more reliable than scrambling because it retains the effects of instrumental behavior with specific temporal patterns.
In conclusion, we believe our 
results give the most reliable 
FAP reported on Kepler data to date, and should be used in validation of
Earth-like candidates at the 
detection threshold.






\matmethods{


To establish the validity of the NST method we here test it on different simulations, which will be chosen to mimic different features in the Kepler data \citep{jon_m_jenkins_kepler_2017}.
Our simulated light-curves consist of $1400$ days of flux measurements which are equally spaced in time with half an hour increments. The gaps in the data match the Kepler-90 star light curve. We perform three experiments, each with a different noise generating process:
\begin{itemize}
    \item \textit{Stellar variability}: this dataset contains white Gaussian observation noise and the stellar variability, which is modeled well by the correlated stationary Gaussian noise \citep{GMF}. For its power spectrum we take the power spectrum of Kepler 90 data, as extracted in \cite{GMF}.
    \item \textit{Rolling bands}: Here we mimic the impact of rolling bands by making the power spectrum of the noise time-dependent, rendering the noise non-stationary. We model the time-series as two interchanging eras, the rolling band and the quiet one, each with its own power spectrum. 
    Time series is a mixture $x(t) = w(t) x_{RB}(t) + (1-w(t) ) x_{\mathrm{quiet}}(t)$, where $x_{RB}(t)$ and $x_{\mathrm{quiet}}(t)$ are stationary time series with the two power spectra and $w(t)$ is the rolling band window. 
    We use power spectra from a star with apparent rolling bands, Kepler ID 7026522. We extract the power spectra by maximizing the likelihood of the light curve model over its parameters: era change times and the two power spectra.
    In simulations we then use a different window function for each simulation. Let's define $\xi(t) \equiv (t- t_0) / P_{RB} \, \mathrm{mod} \, 1$, 
    where $P_{RB} = 372.57$ days is the period of the rolling bands and we randomly select $t_0 \sim \mathcal{U}(0, P_{RB})$ for each simulation.
    The window function $w(t)$ equals 1 inside the rolling band region 
    ($\xi < \Delta_{RB}$),
    0 outside the rolling band region
    ($\Delta_{RB} + \Delta_{T} < \xi < 1 - \Delta_T$) and linearly interpolates between the two in the transition regions ($\Delta_{RB} < \xi < \Delta_{RB} + \Delta_T$ and $1 - \Delta_T < \xi < 1$). The fractional size of the rolling band and transition regions is respectively $\Delta_{RB} = 0.28$ and $\Delta_T = 0.06$.
    \item \textit{Outliers}: On top of the first data-set, we inject non-Gaussian outliers, as in \cite{GMF}. These outliers are uncorrelated and occur at random times, but have a distribution with very long tails.
\end{itemize}
An example realization for each of these experiments is shown in Figure \ref{fig: lc}.

For the test statistic we take a prior corrected likelihood ratio:
\begin{equation} \label{eq: test stat}
    q = \max_{\boldsymbol{z}} \{ SNR(\boldsymbol{z})^2 + 2 \log p(\boldsymbol{z}) / p(\boldsymbol{z}_{\mathrm{ref}}) \},
\end{equation}
which is a simple approximation to the Bayes factor that does not require computing the evidence integral \citep{robnik_statistical_2022}.
Here, $\boldsymbol{z} = (P, \phi, \tau)$ are the parameters of the planet that we maximize over. Planet's period is taken in the range $2 \text{ days} < P < 400 \text{ days}$, phase in range $0 <  \phi < P$ and transit duration $\tau$ in its prior range, as described in \cite{GMF}.
Signal-to-noise-ratio (SNR) is computed by matched filtering with the planet template form \citep{GMF}, using the noise power spectrum, assumed here to be known in advance. $p(\boldsymbol{z})$ is the prior distribution which penalizes the transit durations that are inconsistent with the Kepler's third law and measured stellar density. $\boldsymbol{z}_{\mathrm{ref}}$ are just some reference value of the parameters (taken here to be Earth's parameters) which have no impact on $q$ as a test statistic. Their role is to shift $q$ such that the prior penalty is around zero for reasonable value of parameters.

\subsubsection*{Null dataset}
We start by testing that NST and periodic template false alarm distributions are identical on the null data. We generate 8092 noise realizations for each of the three experiments and search these datasets for planets, both with NST and periodic template. For comparison, we also perform the same analysis on the inverted and scrambled dataset. The
latter is obtained by randomly permuting Kepler's quarters.

The resulting false alarm distributions are shown in Figure \ref{fig: sim0}. 
Inverting the data produces the correct false alarm distribution for the Stellar variability and Rolling bands experiments, but completely fails on the Outliers experiment, because there are significantly more positive than negative outliers, which is not preserved by inversion. 
Scrambling the data produces the correct false alarm distribution for the Stellar variability and Outliers experiments, but fails on the Rolling bands experiment, because it does not preserve the quasi-periodic structure of the noise.
NST method works well in all three experiments.

\subsubsection*{Injected dataset}
Next we inject a periodic planet signature in the data. For each noise realization we draw the period of the injected planet log-uniformly between 2 days and 400 days, the phase uniformly between 0 and $P$ and the transit duration from the transit duration prior \citep{robnik_statistical_2022}. We draw the amplitude of the injected planet uniformly in the range 
3900--5900 for Stellar variability experiment, 
2600--5200 for the Rolling band experiments and
5200--47000 for the Outliers experiment,
all in parts per million. These numbers are selected so that the injected signal can be detected above the noise background.

We then perform the exoplanet search both with the NST and periodic template, the results are shown in Figure \ref{fig: sim1}. The test-statistic distribution for periodic template (green) differs significantly from the distribution on the null data set (black), demonstrating that the periodic template easily picks up the injected signal. NST (orange), on the other hand, does not trigger on the injected signal and reproduces the previous no-signal distribution
without injections. 
Together, these experiments demonstrate that NST can be used as an effective null simulation, regardless of whether the data contains the true signal or not.

\begin{figure*}
    \centering
    \includegraphics[scale=0.35]{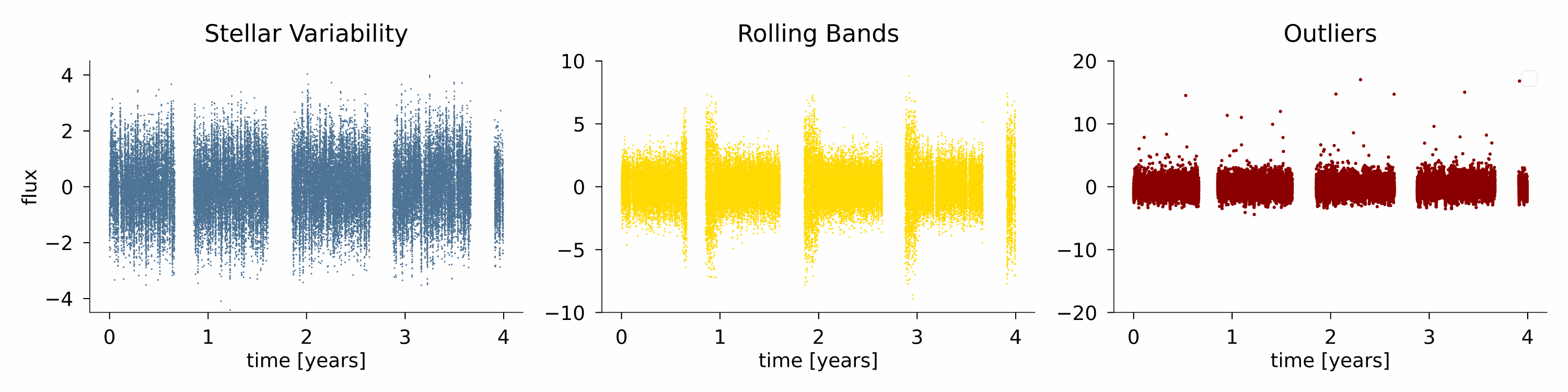}
    \caption{
    Noise realizations for simulated experiments. Stellar variability is a stationary correlated Gaussian noise with a red power spectrum. Rolling bands are a non-stationary correlated Gaussian noise. Periodically the variance increases. Outliers contain stellar variability and uncorrelated non-Gaussian outliers which have a distribution with very long power-law tails.
    }
    \label{fig: lc}
\end{figure*}
\begin{figure*}
    \centering
    \includegraphics[scale=0.35]{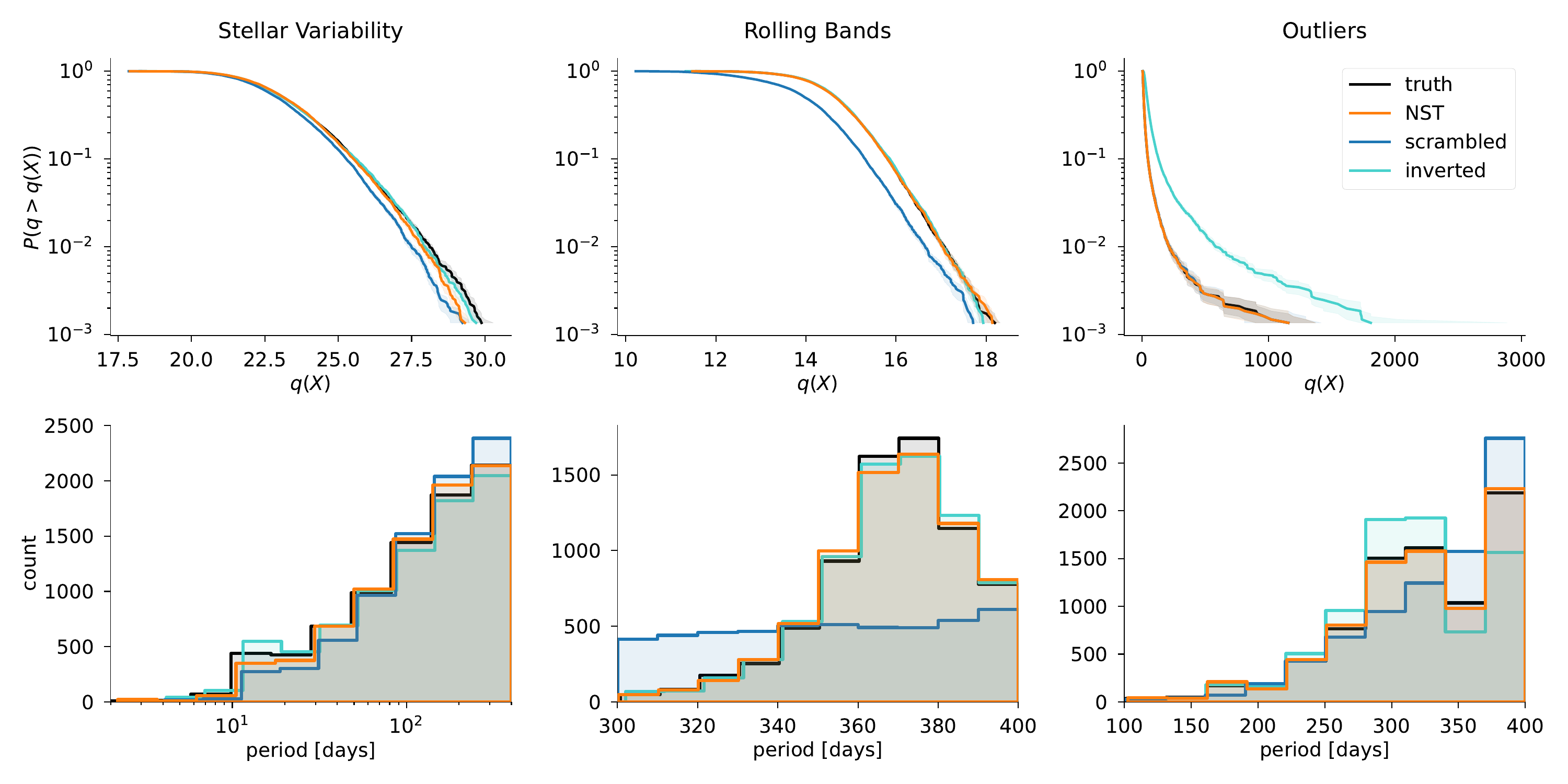}
    \caption{
    Top: Cumulative density distribution of the test scores \eqref{eq: test stat} on the null simulations. The uncertainties are obtained by bootstrap. 
    Bottom: distribution of the best fit (maximizing \eqref{eq: test stat}) period on the null simulations. 
    The periodic template (truth) is shown in black, NST in orange, periodic template on scrambled dataset in blue and periodic template on inverted dataset in cyan. Scrambling does not preserve the quasi-periodicity of the Rolling band noise (middle) and does not produce the correct false alarm distribution, both in terms of the test scores and the inferred periods. Inversion does not produce the correct false alarm distribution on the Outlier dataset (right), since the outlier distribution is not symmetric to inversion. NST distribution is indistinguishable from the truth in all three experiments.
    }
    \label{fig: sim0}
\end{figure*}
\begin{figure*}
    \centering
    \includegraphics[scale=0.35]{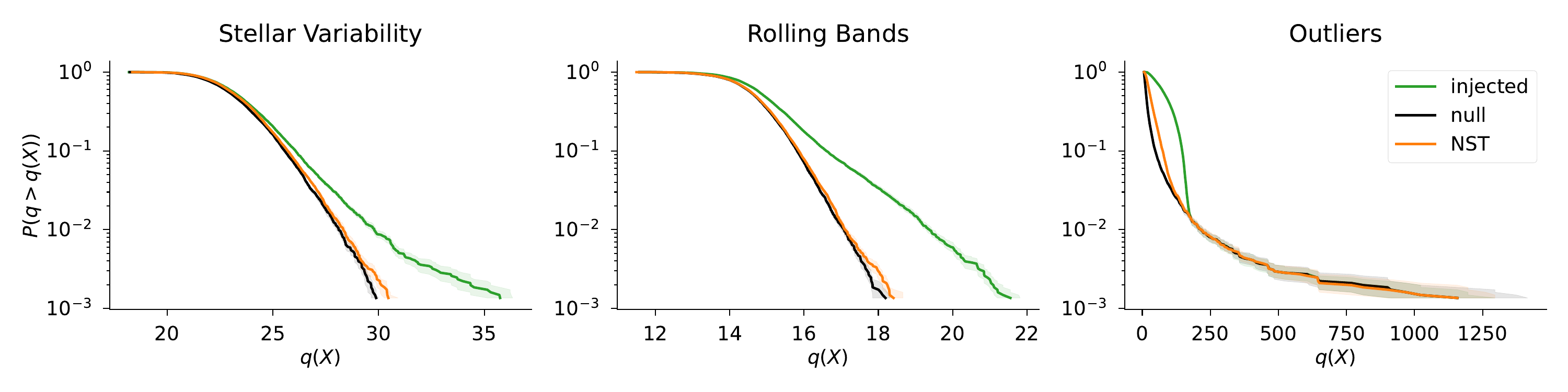}
    \caption{
    Cumulative density distribution of the test statistic from \eqref{eq: test stat} on the null simulations with injected planets. Periodic template distribution significantly shifts if the planets are injected (green vs black), whereas the NST (orange) does not trigger on injections, maintaining the null distribution. Thus NST serves as a null simulation, regardless of the presence of the signal.
    }
    \label{fig: sim1}
\end{figure*}

}


\showmatmethods{} 

    

\begin{table*}[t!]
    \centering
    \caption{Properties of the planets candidates from \citep{bryson_occurrence_2021}, sorted by $\log B$. Their name is printed in bold if they have FAP below $1\%$ as determined by NSTs. $I$ is instellation flux in units of Earth's instellation flux \citep{bryson_occurrence_2021}, $T_{\mathrm{eff}}$ is star effective temperature in Kelvins. Planet's period $P$ is given in days and its radius $R$ in units of Earth's radius. Next we report the number of observed transits for that planet, its previously reported reliability and astrophysical false positive probability (AFPP) from \citep{morton_false_2016, bryson_occurrence_2021, bryson_probabilistic_2020}. Finally we report its $\log B$ and $SNR$ test statistics along with FAP=5\% and 1\% NST quantiles. 
    } \label{table}
    \begin{tabular}{lrrrrrrrrrrrrrr} 
    \bfseries name & \bfseries I [$I_{\oplus}$] & \bfseries $T_{\mathrm{eff}}$ [K] & \bfseries $P$ [days] & \bfseries $R$ [$R_{\oplus}$] & \bfseries $\#$ transits & \bfseries reliability & \bfseries AFPP & \bfseries log B & $5\%$ & $1\%$ & \bfseries SNR & $5\%$ & $1\%$ \\ 
    \midrule
    \textbf{Kepler 174 d} & $0.56^{+0.04}_{-0.04}$ & $4918^{+90}_{-88}$ & 247.353 & $2.45^{+0.11}_{-0.07}$ & 5 & 1.0 & 0.0 & 660.6 & 15.7 & 17.7 & 37.1 & 8.7 & 8.9 \\
    \textbf{Kepler 62 e} & $1.44^{+0.11}_{-0.1}$ & $4966^{+82}_{-82}$ & 122.385 & $1.8^{+0.07}_{-0.04}$ & 11 & 1.0 & 0.0 & 583.2 & 18.2 & 20.2 & 34.9 & 9.0 & 9.3 \\
    \textbf{Kepler 26 e} & $2.0^{+0.17}_{-0.17}$ & $4124^{+43}_{-68}$ & 46.827 & $2.36^{+0.09}_{-0.13}$ & 30 & 1.0 & 0.0 & 347.8 & 22.5 & 24.7 & 27.4 & 9.5 & 9.8 \\
    \textbf{Kepler 22 b} & $0.96^{+0.07}_{-0.07}$ & $5625^{+93}_{-93}$ & 289.862 & $2.22^{+0.1}_{-0.3}$ & 4 & 1.0 & 0.04 & 326.5 & 30.2 & 35.0 & 26.6 & 10.4 & 11.1 \\
    \textbf{Kepler 1539 b} & $1.6^{+0.15}_{-0.15}$ & $5051^{+90}_{-85}$ & 133.303 & $2.42^{+0.14}_{-0.11}$ & 10 & 1.0 & 0.0 & 224.7 & 18.6 & 25.4 & 22.4 & 9.2 & 9.7 \\
    \textbf{Kepler 1816 b} & $2.01^{+0.14}_{-0.14}$ & $4944^{+75}_{-74}$ & 91.501 & $1.77^{+0.1}_{-0.05}$ & 14 & 1.0 & 0.0 & 198.8 & 23.3 & 30.0 & 21.2 & 9.7 & 10.5 \\
    \textbf{KOI 2770.01} & $0.45^{+0.04}_{-0.03}$ & $4475^{+80}_{-75}$ & 205.385 & $2.26^{+0.13}_{-0.08}$ & 7 & 1.0 & 0.01 & 190.0 & 19.2 & 21.6 & 20.8 & 9.0 & 9.2 \\
    \textbf{Kepler 439 b} & $1.97^{+0.16}_{-0.15}$ & $5545^{+94}_{-94}$ & 178.14 & $2.36^{+0.14}_{-0.09}$ & 8 & 1.0 & 0.01 & 170.8 & 14.5 & 17.3 & 19.8 & 8.6 & 8.9 \\
    \textbf{Kepler 1040 b} & $2.1^{+0.15}_{-0.15}$ & $5756^{+97}_{-96}$ & 201.119 & $2.34^{+0.09}_{-0.06}$ & 7 & 1.0 & 0.0 & 147.1 & 21.7 & 23.3 & 18.6 & 9.4 & 9.6 \\
    \textbf{Kepler 1362 b} & $1.09^{+0.11}_{-0.1}$ & $4775^{+91}_{-83}$ & 136.205 & $2.48^{+0.08}_{-0.2}$ & 10 & 1.0 & 0.0 & 137.5 & 18.2 & 19.6 & 18.0 & 8.8 & 8.9 \\
    \textbf{Kepler 1881 b} & $1.65^{+0.2}_{-0.18}$ & $5397^{+103}_{-100}$ & 159.389 & $2.48^{+0.16}_{-0.13}$ & 9 & 1.0 & 0.01 & 122.1 & 20.9 & 24.7 & 17.1 & 9.2 & 9.6 \\
    \textbf{Kepler 1544 b} & $0.77^{+0.05}_{-0.05}$ & $4697^{+76}_{-68}$ & 168.811 & $1.71^{+0.12}_{-0.08}$ & 9 & 1.0 & 0.0 & 117.8 & 21.7 & 23.8 & 16.8 & 9.3 & 9.6 \\
    \textbf{Kepler 1701 b} & $1.34^{+0.11}_{-0.1}$ & $5216^{+91}_{-86}$ & 169.133 & $2.22^{+0.35}_{-0.23}$ & 9 & 1.0 & 0.0 & 112.0 & 24.0 & 29.0 & 16.5 & 9.7 & 10.4 \\
    \textbf{Kepler 440 b} & $0.82^{+0.07}_{-0.07}$ & $4171^{+56}_{-49}$ & 101.11 & $1.8^{+0.1}_{-0.08}$ & 10 & 1.0 & 0.0 & 99.6 & 24.6 & 30.4 & 15.9 & 9.6 & 10.3 \\
    \textbf{Kepler 1389 b} & $1.79^{+0.14}_{-0.13}$ & $4806^{+84}_{-76}$ & 99.253 & $2.22^{+0.05}_{-0.59}$ & 15 & 1.0 & 0.01 & 96.8 & 20.5 & 23.5 & 15.5 & 9.2 & 9.6 \\
    \textbf{Kepler 1455 b} & $1.66^{+0.16}_{-0.15}$ & $4050^{+64}_{-69}$ & 49.277 & $1.89^{+0.11}_{-0.1}$ & 25 & 1.0 & 0.0 & 94.0 & 22.2 & 26.8 & 15.5 & 9.7 & 10.1 \\
    \textbf{Kepler 1549 b} & $1.14^{+0.1}_{-0.09}$ & $5288^{+94}_{-89}$ & 214.886 & $2.32^{+0.16}_{-0.09}$ & 7 & 0.99 & 0.0 & 76.1 & 18.0 & 18.7 & 14.1 & 8.9 & 9.5 \\
    \textbf{Kepler 442 b} & $1.01^{+0.08}_{-0.07}$ & $4602^{+84}_{-76}$ & 112.302 & $1.35^{+0.08}_{-0.08}$ & 12 & 1.0 & 0.09 & 61.1 & 20.6 & 29.8 & 12.9 & 9.3 & 10.1 \\
    \textbf{Kepler 1126 c} & $2.09^{+0.19}_{-0.18}$ & $5814^{+116}_{-112}$ & 199.662 & $1.42^{+0.09}_{-0.07}$ & 5 & 0.99 & 0.0 & 57.0 & 16.9 & 19.3 & 12.6 & 8.9 & 9.1 \\
    \textbf{Kepler 351 d} & $1.72^{+0.21}_{-0.19}$ & $5213^{+97}_{-91}$ & 142.54 & $2.4^{+0.17}_{-0.13}$ & 8 & 1.0 & 0.0 & 54.7 & 16.3 & 18.0 & 12.5 & 8.7 & 9.0 \\
    \textbf{Kepler 1606 b} & $1.29^{+0.11}_{-0.11}$ & $5361^{+91}_{-89}$ & 196.436 & $2.06^{+0.14}_{-0.1}$ & 7 & 0.99 & 0.0 & 54.3 & 16.5 & 23.9 & 12.4 & 8.9 & 9.4 \\
    \textbf{KOI 2719.02} & $1.25^{+0.1}_{-0.09}$ & $4601^{+81}_{-76}$ & 106.26 & $1.25^{+0.15}_{-0.08}$ & 14 & 1.0 & 0.04 & 45.2 & 22.2 & 24.2 & 11.7 & 9.4 & 9.6 \\
    \textbf{KOI 7345.01} & $1.18^{+0.12}_{-0.11}$ & $5883^{+113}_{-111}$ & 377.499 & $2.44^{+0.19}_{-0.13}$ & 4 & 0.89 & 0.01 & 44.5 & 12.9 & 14.4 & 11.7 & 8.6 & 8.9 \\
    \textbf{Kepler 443 b} & $0.78^{+0.08}_{-0.07}$ & $4790^{+84}_{-78}$ & 177.668 & $2.37^{+0.51}_{-0.33}$ & 8 & 0.99 & 0.0 & 31.8 & 22.0 & 24.9 & 10.5 & 9.5 & 9.7 \\
    \textbf{Kepler 441 b} & $0.24^{+0.02}_{-0.02}$ & $4147^{+67}_{-45}$ & 207.248 & $1.48^{+0.09}_{-0.11}$ & 6 & 0.99 & 0.0 & 29.3 & 19.9 & 25.4 & 10.3 & 9.3 & 9.8 \\
    \textbf{KOI 2184.02} & $1.66^{+0.14}_{-0.13}$ & $4820^{+88}_{-83}$ & 95.906 & $1.93^{+0.05}_{-0.21}$ & 14 & 1.0 & 0.03 & 27.2 & 18.1 & 20.6 & 10.1 & 9.1 & 9.4 \\
    Kepler 186 f & $0.4^{+0.03}_{-0.03}$ & $4023^{+58}_{-62}$ & 129.943 & $1.43^{+0.14}_{-0.25}$ & 10 & 0.98 & 0.06 & 26.8 & 31.7 & 34.2 & 10.0 & 10.5 & 10.7 \\
    \textbf{KOI 2194.03} & $1.36^{+0.13}_{-0.12}$ & $5965^{+122}_{-116}$ & 445.198 & $1.8^{+0.1}_{-0.14}$ & 4 & 0.71 & 0.05 & 25.6 & 16.4 & 18.1 & 9.6 & 9.0 & 9.2 \\
    KOI 5433.01 & $1.81^{+0.19}_{-0.18}$ & $5798^{+112}_{-110}$ & 237.816 & $2.42^{+0.16}_{-0.13}$ & 6 & 0.98 & 0.0 & 24.5 & 26.1 & 28.6 & 9.8 & 9.9 & 10.2 \\
    \textbf{KOI 3344.03} & $1.44^{+0.14}_{-0.14}$ & $5495^{+96}_{-95}$ & 208.543 & $2.13^{+0.15}_{-0.16}$ & 4 & 0.97 & 0.0 & 24.0 & 17.7 & 19.6 & 9.5 & 8.9 & 9.1 \\
    \textbf{Kepler 452 b} & $1.11^{+0.08}_{-0.08}$ & $5900^{+102}_{-100}$ & 384.842 & $1.46^{+0.09}_{-0.08}$ & 4 & 0.68 & 0.01 & 21.2 & 17.1 & 19.5 & 9.4 & 9.1 & 9.3 \\
    KOI 8159.02 & $1.8^{+0.15}_{-0.15}$ & $6290^{+121}_{-118}$ & 353.017 & $2.2^{+0.13}_{-0.1}$ & 3 & 0.85 & 0.0 & 20.2 & 37.7 & 41.7 & 9.4 & 11.0 & 11.4 \\
    KOI 6971.01 & $1.23^{+0.1}_{-0.09}$ & $4921^{+82}_{-83}$ & 129.218 & $1.69^{+0.27}_{-0.31}$ & 10 & 0.99 & 0.01 & 15.3 & 24.7 & 37.2 & 8.6 & 9.7 & 11.0 \\
    KOI 7882.01 & $1.87^{+0.15}_{-0.13}$ & $4390^{+81}_{-74}$ & 65.415 & $1.33^{+0.08}_{-0.15}$ & 21 & 0.98 & 0.08 & 14.2 & 16.6 & 19.1 & 8.5 & 8.8 & 9.1 \\
    KOI 7931.01 & $1.61^{+0.16}_{-0.15}$ & $5843^{+106}_{-100}$ & 242.065 & $1.75^{+0.1}_{-0.19}$ & 6 & 0.89 & 0.06 & 7.6 & 19.7 & 22.9 & 7.4 & 9.4 & 9.8 \\
    KOI 8048.01 & $1.31^{+0.11}_{-0.11}$ & $6058^{+108}_{-107}$ & 379.67 & $1.76^{+0.16}_{-0.15}$ & 3 & 0.48 & 0.01 & 6.9 & 17.6 & 22.6 & 7.4 & 9.0 & 9.3 \\
    KOI 7746.01 & $1.02^{+0.15}_{-0.13}$ & $6135^{+118}_{-114}$ & 196.999 & $2.15^{+0.12}_{-0.26}$ & 7 & 0.6 & 0.06 & 3.1 & 11.7 & 13.5 & 7.2 & 8.4 & 8.7 \\
    KOI 8047.01 & $0.38^{+0.03}_{-0.03}$ & $4712^{+78}_{-74}$ & 302.331 & $1.86^{+0.14}_{-0.22}$ & 3 & 0.84 & 0.14 & -0.3 & 18.9 & 20.4 & 6.7 & 9.0 & 9.1 \\
    KOI 8246.01 & $1.7^{+0.16}_{-0.15}$ & $6091^{+125}_{-123}$ & 425.647 & $1.72^{+0.12}_{-0.22}$ & 4 & 0.36 & 0.01 & -1.1 & 21.3 & 23.5 & 6.6 & 9.3 & 9.6 \\
    \bottomrule
    \end{tabular}
\end{table*}

\acknow{
This material is based upon work supported in part by the Heising-Simons Foundation grant 2021-3282. We thank Stephen Bryson for numerous discussions.  
Kepler was competitively selected as the 10th Discovery mission and was funded by NASA's Science Mission Directorate. The authors acknowledge the eﬀorts of the Kepler Mission team in obtaining the light curve data and data validation products used in this publication. These data were generated by the Kepler Mission science pipeline through the eﬀorts of the Kepler Science Operations Center and Science Oﬃce. The Kepler light curves are available at the Mikulski Archive for Space Telescopes, and the Data Validation products are available at the NASA Exoplanet Science Institute.
}

\showacknow{} 


\bibliography{citations,references,referencesReviewer}

\end{document}